\documentstyle[tighten,preprint,aps,epsf]{revtex}
\preprint{KUNS 1319}
\draft
\title{Proton inelastic scattering to continuum studied with \\
 antisymmetrized molecular dynamics}
\author{Eiji I. Tanaka, Akira Ono, Hisashi Horiuchi,
  Tomoyuki Maruyama\thanks{Present address: Advanced Science Research
  Center, Japan Atomic Energy Research Institute, Tokai-mura,
  Ibaraki-ken 319-11, Japan.}, and Andreas Engel}
\address{Department of Physics, Kyoto University, Kyoto 606-01,
Japan}
\begin{document}
\maketitle
\begin{abstract}
Intermediate energy (p, p$'$x) reaction is studied with
antisymmetrized molecular dynamics (AMD) in the cases of $^{58}$Ni
target with $E_p = 120$ MeV and $^{12}$C target with $E_p = $ 200 and
90 MeV.  Angular distributions for various $E_{p'}$ energies are shown
to be reproduced well without any adjustable parameter, which shows
the reliability and usefulness of AMD in describing light-ion
reactions.  Detailed analyses of the calculations are made in the case
of $^{58}$Ni target and following results are obtained: Two-step
contributions are found to be dominant in some large angle region and
to be indispensable for the reproduction of data.  Furthermore the
reproduction of data in the large angle region $\theta \agt 120^\circ$
for $E_{p'}$ = 100 MeV is shown to be due to three-step contributions.
Angular distributions for $E_{p'} \agt$ 40 MeV are found to be
insensitive to the choice of different in-medium nucleon-nucleon cross
sections $\sigma_{NN}$ and the reason of this insensitivity is
discussed in detail.  On the other hand, the total reaction cross
section and the cross section of evaporated protons are found to be
sensitive to $\sigma_{NN}$. In the course of the analyses of the
calculations, comparison is made with the distorted wave approach.
\end{abstract}
\pacs{25.40.Ep, 24.50.+g, 24.10.-i}

\narrowtext
\section{Introduction}

Nucleon inelastic scattering to continuum at intermediate energies has
been extensively studied, through which our understanding of
pre-equilibrium processes in the scattering has much advanced. Among
many kinds of theoretical investigations including the exciton model
\cite{EXCITON,IWAMOTO} and other multi-step reaction theories
\cite{FKK,TUL,NWY}, the intra-nuclear cascade model (INC)
\cite{KIKA,GOLD} has served as an important
approach since the activation of many degrees of freedom in
pre-equilibrium processes is complicated compared to the
compound-nucleus reaction process with full equilibrization.  Because
the original INC model has several drawbacks such as the absence of
the attractive interaction among nucleons, many kinds of modifications
have been introduced into the INC model.

Recently, besides INC and its modified versions, new types of
transport theories (or microscopic simulation theories) like BUU
(Boltzmann-Uehling-Uhlenbeck) \cite{BUU}, QMD (quantum molecular
dynamics) \cite{QMD,BOAL}, and AMD (antisymmetrized molecular
dynamics) \cite{HORI,ONO} have been developed in order to investigate
complicated reaction processes of heavy-ion collisions at intermediate
and high energies.  These new types of transport theories can describe
the self-consistent mean field of the system which changes with time
depending on the stage of the reaction process.  These theories are,
of course, also applicable to light-ion reactions including the
nucleon inelastic scattering to continuum.

We think that QMD and AMD approaches to reactions induced by light
ions and also hadrons and leptons are especially important. These
molecular dynamics models can describe dynamical production processes
of fragments and hence they can provide us with a unified theoretical
description of two different kinds of reaction processes, namely
processes of the type of light-ion physics and those of the type of
heavy-ion physics.  Nucleon inelastic scattering to continuum belongs
to the former type and fragmentation reaction to the latter type.

In this paper we study proton inelastic scattering to continuum at
intermediate energies by the use of the AMD model.  The AMD model is a
new transport theory and has already proved to be very successful in
describing heavy-ion collisions at medium energies
\cite{ONO,ONOA,ONOB,ONOC,BODR,TAOR,VARR}.  AMD describes the total
system with a Slater determinant of nucleon wave packets and hence it
has quantum mechanical character, which has been demonstrated in the
ability of treating shell effects in the dynamical formation of
fragments.  Furthermore it has been shown that ground-state wave
functions of colliding nuclei given by the AMD model are realistic and
reproduce many spectroscopic data very well
\cite{HORIA,NEKANADA,SANTO}.  We report in this paper the study of (p,
p$'$x) reaction in the cases of $^{58}$Ni target with $E_p = 120$ MeV
and $^{12}$C target with $E_p = $ 200 and 90 MeV.  We show that the
angular distributions are well reproduced by AMD without any
adjustable parameter.  This shows the reliability and usefulness of
AMD in treating light-ion reactions.  We make detailed analyses of the
calculations in the case of $^{58}$Ni target in the following way:
We decompose the calculated cross sections into the contributions
coming from different steps in order to study magnitudes of multi-step
contributions.  Two-step contributions are shown to be dominant in
some large angle region and to be indispensable for the reproduction
of data. Furthermore the reproduction of data in the large angle
region $\theta \agt 120^\circ$ for $E_{p'}$ = 100 MeV is shown to be
due to three-step contributions.  Angular distributions for $E_{p'}
\agt$ 40 MeV are shown to be insensitive to the choice of different
in-medium nucleon-nucleon cross sections $\sigma_{NN}$ and the reason
of this insensitivity is discussed in detail.  On the other hand, the
total reaction cross section and cross section of evaporated protons
are shown to be sensitive to $\sigma_{NN}$. In discussing the
calculated results we make comparison with the results obtained with
the semi-classical distorted-wave approach of Ref.\ \cite{KAWAI}.

The organization of this paper is as follows: In Sec.\ II we explain
the AMD framework, the adopted effective two-nucleon force, and three
choices of in-medium nucleon-nucleon cross section $\sigma_{NN}$. A
detailed explanation of the definition of the step number of reaction
process is also given. In Sec.\ III we give the comparison with
experiments of the calculated angular distributions at various
$E_{p'}$ energies. Here the decomposition of the cross sections into
multi-step contributions is also made.  In Sec.\ IV we make detailed
analysis of the dependence of the calculation on the different choice
of $\sigma_{NN}$.  Finally in Sec.\ V we give the summary.

\section{Formulation}
\subsection{AMD}

The framework of AMD (antisymmetrized molecular dynamics) was
described in detail in Ref.\ \cite{ONO} and hence we here explain only
the outline of the AMD theory.

In AMD, the wave function of $A$-nucleon system is described by a
Slater determinant $|\Phi(Z)\rangle$,
\begin{equation}
|\Phi(Z)\rangle = {1\over\sqrt{A!}}\det
\Bigl[\varphi_j(k)\Bigr], \quad
\varphi_j = \phi_{{\bf Z}_j} \chi_{{\alpha}_j}
\label{AMDWF}
\end{equation}
where $\chi$ stands for the spin-isospin function and $\alpha_j$
represents the spin-isospin label of the $j$-th single particle state,
$\alpha_j={\rm p}\uparrow$, ${\rm p}\downarrow$, ${\rm n}\uparrow$, or
${\rm n}\downarrow$.  $\phi_{{\bf Z}_j}$ is the spatial wave function
of the $j$-th single-particle state which is a Gaussian wave packet,
\begin{equation}
\begin{array}{rcl}
 \langle {\bf r} | \phi_{{\bf Z}_j} \rangle & = &
\displaystyle
 \Bigl( {2\nu \over \pi} \Bigr)^{3/4} \exp \Bigl[ -\nu \Bigl(
 {\bf r} - {{\bf Z}_j \over {\sqrt \nu}} \Bigr)^2 +
      {1 \over 2}{\bf Z}^2_j \Bigr], \\
 {\bf Z}_j  & = & \displaystyle{\sqrt{\nu} {\bf D}_j +
   {i \over 2\hbar\sqrt{\nu}} {\bf K}_j},
\end{array}
\label{GAUSSWF}
\end{equation}
where the width parameter $\nu$ is treated as time-independent in the
present work.  We take $\nu$=0.16 fm$^{-2}$ in the calculation in this
paper.  Here ${\bf Z}_j$ is the complex vector whose real and
imaginary parts, ${\bf D}_j$ and ${\bf K}_j$, are the spatial and
momentum centers of the packet, respectively.

The time developments of the coordinate parameters, $Z = \lbrace {\bf
Z}_j \ ( j=1,2, \ldots, A )\rbrace$, are determined by two processes.
One is the time development determined by the time-dependent
variation-al principle;
\begin{eqnarray}
  \delta \int_{t_1}^{t_2} dt
  { \langle \Phi(Z) | \Bigl(
  i\hbar {\displaystyle {d\over dt}}-H \Bigr) |
  \Phi(Z) \rangle  \over \langle \Phi(Z) |
  \Phi(Z) \rangle } = 0,
\label{eq}
\end{eqnarray}
which leads to the equation of motion for $Z$,
\begin{equation}
\begin{array}{rcl}
&& \displaystyle{
i\hbar \sum_{j\tau} C_{k\sigma,j\tau} {d\over dt} Z_{j\tau} =
  {\partial \over \partial {Z_{k\sigma}^*}}
  { \langle \Phi(Z) | H | \Phi(Z) \rangle
  \over \langle \Phi(Z) | \Phi(Z) \rangle}}, \\
&& \displaystyle{
C_{k\sigma,j\tau}
  \equiv {\partial^2\over\partial Z_{k\sigma}^*\partial Z_{j\tau}}
      \ln \langle \Phi(Z) | \Phi(Z) \rangle} ,
\end{array}
\label{EQMTN}
\end{equation}
where $\sigma, \tau=x,y,z$.

During the dynamical reaction stage, the total system can be separated
into several isolated nucleons and fragments.  Since the wave
functions of the center-of-mass motion of these isolated nucleons and
fragments are Gaussian wave packets, each of these isolated particles
carries spurious zero-point energy of its center-of-mass motion.  The
total amount of the spurious energy of center-of-mass motion can be
expressed as a function of $Z$ \cite{ONO,ONOB}, which we denote as
$E_{\rm sprs}(Z)$.  The actual Hamiltonian we use in the above
equation of motion (Eq.\ (\ref{EQMTN})) is, therefore, given by
$\langle \Phi(Z) | H | \Phi(Z) \rangle /\langle \Phi(Z) | \Phi(Z)
\rangle - E_{\rm sprs}(Z)$.

The second process which determines the time development of the system
is the stochastic two-nucleon collision process.  We incorporate this
process in a similar way as it is done in QMD by introducing the
physical nucleon coordinates $\{ {\bf W}_j \}$, mimicking the
time-dependent cluster model (TDCM) \cite{SARA}, as
\begin{equation}
\begin{array}{rcl}
  {\bf W}_j &=& \displaystyle
\sqrt\nu{\bf R}_j+{i\over2\hbar\sqrt\nu}{\bf P}_j
        \equiv  \sum_{k=1}^A \Bigl(\sqrt Q\Bigr)_{jk}{\bf Z}_k, \\
  Q_{jk} & \equiv &
  \displaystyle{\partial\over\partial({\bf Z}_j^*\cdot{\bf Z}_k)}
         \ln \langle \Phi(Z)| \Phi(Z) \rangle.
\end{array}
\label{DefOfW}
\end{equation}
Here it should be noted that, due to the antisymmetrization, ${\bf
D}_j$ and ${\bf K}_j$ of ${\bf Z}_j$ do not represent the position and
momentum of $j$-th nucleon, respectively. When physical nucleon
positions ${\bf R}_j$ and ${\bf R}_k$ become close each other, these
$j$-th and $k$-th nucleons can make stochasic two-nucleon
collisions. Let initial ${\bf W}_j$ and ${\bf W}_k$ be changed into
final ${\bf W}'_j$ and ${\bf W}'_k$ by a two-nucleon collision.  In
order to continue the calculation of time development of the system
wave function after this collision, we need to back-transform $\{{\bf
W}_1,\dots,{\bf W}'_j,
\ldots,{\bf W}'_k,\ldots,{\bf W}_A\}$ into $\{{\bf Z}'_1,{\bf Z}'_2,
\ldots,{\bf Z}'_A\}$.
However, in general, the back-transformation from $W =
\{{\bf W}_j \ (j=1,\ldots,A) \}$ to $Z =
\{{\bf Z}_j \ (j=1,\ldots,A) \}$ does not always exist.  When the
back-transformation does not exist, we regard that the two-nucleon
collision is Pauli-blocked. $W$ is defined to be in
Pauli-forbidden region if it cannot be back-transformed to any $Z$.
The notion of the Pauli-forbidden region defined above is an
extension of that of TDCM \cite{SARA}. AMD without stochastic
two-nucleon collisions is the same as FMD (Fermionic Molecular
Dynamics) \cite{FELD}, and FMD is a special case of TDCM
\cite{TDCM,SARA} where every cluster is composed of a single nucleon.

The full procedure of the AMD description of the nuclear reaction
consists of three major steps: The first step is the initialization,
namely the construction of the wave functions of the ground states of
colliding nuclei.  The initialization is made by the use of the
frictional cooling method \cite{HORI,HORIA,NEKANADA,SANTO}.  It has
been checked that wave functions given by AMD are realistic and
reproduce many spectroscopic data very well. The second step is the
calculation of dynamical collision stage by the equation of motion and
stochastic two-nucleon collisions. The final step is the calculation
of the statistical decay of primordial fragments.  Primordial
fragments mean the fragments which are present when the dynamical
stage of the reaction has finished.  These fragments are not in their
ground states but are excited, and they decay through evaporation with
a long time scale.  In this paper, the switching time from the
dynamical stage to the evaporation stage was chosen to be 150 fm/c.
Statistical decays of fragments were calculated with the code of Ref.\
\cite{MARU} which is similar to the code of P\"uhlhofer \cite{PUHL}.

\subsection{Effective force and in-medium nucleon-nucleon
cross sections}

As the effective two-nucleon force, we adopt the Gogny force
\cite{GOGNY} which has been successfully used in studying heavy-ion
reactions with AMD \cite{ONOB,ONOC}.  The Gogny force consists of
finite-range two-body force and density-dependent zero-range repulsive
force. This force gives a momentum-dependent mean field which
reproduces well the observed energy dependence of the nucleon optical
potential up to about 200 MeV.  The nuclear matter EOS given by this
force is soft with the incompressibility $K = $ 228 MeV. Corresponding
to the choice of Gogny force, the calculational formula of the total
spurious center-of-mass energy $E_{\rm sprs}(Z)$ is taken to be the
same as Ref.\ \cite{ONOB} in the case of $^{12}$C target, while in the
case of $^{58}$Ni target the value of the $T_0$ parameter in the
formula of $E_{\rm sprs}(Z)$ is changed into 8.70 MeV leaving other
parameters unchanged.  The binding energies of $^{12}$C and $^{58}$Ni
are calculated to be 92.6 MeV and 507.6 MeV, respectively, while the
observed values are 92.2 MeV and 506.5 MeV, respectively.  The
r.m.s. radii of $^{12}$C and $^{58}$Ni are calculated to be 2.55 fm
and 3.85 fm, respectively, which are reasonable.

As the in-medium nucleon-nucleon cross section $\sigma_{NN}$, we adopt
the following three different ones, case-1, case-2, and case-3:
Case-1 $\sigma_{NN}$ including its angular distribution is the same as
the $\sigma_{NN}$ used in Ref.\ \cite{ONOB} and is given as
$\sigma_{NN} = \mathop{\rm min}\{\sigma_{NN}^H, \sigma_{NN}^L \}$.
Here $\sigma_{NN}^H$ defines the cross section for the high energy
region and is the same as
the free cross section which is parametrized as
\begin{equation}
\begin{array}{rcl}
\sigma_{pn}^H & = & \max \Bigl\{ 13335 (E /{\rm MeV})^{-1.125},
40 \Bigr\} \  {\rm mb}, \\
\sigma_{pp}^H & = & \sigma_{nn}^H =
  \max\Bigl\{ 4445 (E /{\rm MeV})^{-1.125}, 25 \Bigr\} \,{\rm mb}.
\label{SIGMAH}
\end{array}
\end{equation}
where $E$ is the laboratory energy.  On the other hand,
$\sigma_{NN}^L$ defines the cross section for the low energy region.
It is density-dependent and is given as
\begin{eqnarray}
\sigma_{pn}^L & = & \sigma_{pp}^L  = \sigma_{nn}^L \nonumber\\
& = & {100 \  {\rm mb} \over 1 + E/(200 {\rm MeV}) +
C \mathop{\rm min} {\displaystyle \{} (\rho/\rho_0)^{1/2}, 1
{\displaystyle \}} },\label{SIGMAL}\\
C & = & 2. \nonumber
\end{eqnarray}
where $\rho_0$ is the normal density, $\rho_0 = 0.17\ {\rm
fm}^{-3}$. The center-of-mass angular distribution of $pp$ and $nn$
collisions is taken to be isotropic while that of $pn$ collisions is
given as
\begin{equation}
\begin{array}{rcl}
&& \displaystyle{
{d\sigma_{pn} \over d\Omega} \propto
10^{-\alpha {\displaystyle (} \pi/2 - |\theta - \pi/2|
{\displaystyle )} }}, \\
&& \displaystyle{
\alpha  = {2 \over \pi} \max
\Bigl\{ 0.333 \ln (E / (1{\rm MeV})) -1, 0 \Bigr\}.}
\label{SIGMAANG}
\end{array}
\end{equation}

Case-2 $\sigma_{NN}$ is the same as case-1 $\sigma_{NN}$
except that $C$ of $\sigma_{NN}^L$ is taken to be $C = 0$ instead of
$C = 2$.  Hence case-2 $\sigma_{NN}$ is not dependent on density.
Case-3 $\sigma_{NN}$ including its angular distribution is the
same as the cross section adopted in Ref.\ \cite{KAWAI}.  It is just
the free cross section and its parametrization is taken from Ref.\
\cite{KIKA} as follows;
\begin{eqnarray}
\sigma_{pn} & = & \Bigl( {34.10 \over \beta^2} - {82.2 \over \beta}
+ 82.2 \Bigr)  \  {\rm mb}, \nonumber\\
\sigma_{pp} & = & \sigma_{nn} =  \Bigl( {10.63 \over \beta^2} -
{29.92 \over \beta} + 49.9 \Bigr)  \  {\rm mb},
\label{SIGMAKK} \\
  \beta & = & {v \over c} = \sqrt{2E \over mc^2} .\nonumber
\end{eqnarray}
The center-of-mass angular distribution is taken from Ref.\
\cite{BERTINI}, where $pp$ and $nn$ collisions is isotropic
while that of $pn$ collisions for 40 MeV $< E < 280$ MeV is given as
\begin{equation}
{d\sigma_{pn} \over d\Omega} \propto \left\{
\begin{array}{rl}
 A_1(E) + B_1(E) (\cos \theta)^3, \quad
        &\mbox{for $0 \leq \theta \leq \pi/2$}\\
        A_1(E) + B_2(E) (\cos \theta)^4, \quad
        &\mbox{for $\pi/2 \leq \theta \leq \pi$}
\end{array}\right.
\label{SIGMAANGZ}
\end{equation}
with values of $A_1(E), B_1(E)$, and $B_2(E)$ being shown in Table I.
The $pn$ collisions below 40 MeV is taken to be isotropic.

In this paper we study $^{12}$C(p, p$'$x) by using only case-1
$\sigma_{NN}$, while we apply all the three kinds of $\sigma_{NN}$ to
the study of $^{58}$Ni(p, p$'$x).

\subsection{Step number of reaction process}

Here for the sake of later discussion of multi-step processes, we
define the number of steps in our AMD approach.  What we need is to
determine for each out-coming proton the step number of the reaction
process. The step number should be defined as the number of
collisions which have contributed in emitting the nucleon. If an
outcoming proton originates from the decay of some primordial
fragment, the process which this proton has experienced is a
compound-nucleus process and we need not to define the step number;
namely the step number is concerned only with protons which are
emitted dynamically.

The step numbers are defined and calculated in the following way. We
put a label to every nucleon so that the label of the incident proton
is 1.  Let us consider the first collision of the incident proton with
a target nucleon with label $k$.  After the first collision, each of
the two nucleons 1 and $k$ may further make collisions with other
target nucleons. We put ordering numbers to all these successive
collisions beginning with the first collision.  We call the ordering
number of collision the collision index.  The collision index of the
first collision is No.1.  Just after the first collision, we let each
nucleon $i$ of the total system carry its respective set $C_i(1)$
composed of related previous collision indices.  The sets $C_1(1)$ and
$C_k(1)$ of the two nucleons 1 and $k$ are $C_1(1) = C_k(1) = \{ {\rm
No.}1 \}$ but the sets $C_i(1)$ of other nucleons (\ $i \neq 1, k$\ )
are all null, namely $C_i(1) = \emptyset$ ($i \neq 1, k$).  One of
the nucleons 1 and $k$ can make the collision with collision index
No.2.  Let us consider the case that the nucleon $k$ makes the No.2
collision with a nucleon $j$.  Then, just after the No.2 collision we
let the nucleons $k$ and $j$ carry not the old sets $C_k(1) = \{ {\rm
No.}1 \}$ and $C_j(1) = \emptyset$ but new sets $C_k(2) = C_j(2) = \{
{\rm No.}1, {\rm No.}2 \}$.  However, the sets $C_i(2)$ of other
nucleons than $k$ and $j$ after the No.2 collision are made unchanged,
namely $C_i(2) = C_i(1)$ for $i \neq k, j$. In general, if a nucleon
$m$ makes the No.$p$ collision with a nucleon $n$, the new sets after
the No.$p$ collision are
\begin{equation}
\begin{array}{rcl}
&& C_m(p) = C_n(p) = C_m(p-1) \cup C_n(p-1) \cup \{ {\rm No.}p \}, \\
\label{STEPNUMBER}
&& C_i(p) = C_i(p-1) \quad\mbox{for $i \neq m, n$}.
\end{array}
\end{equation}
Note that the double counting of the same collision index is to be
avoided in constructing new sets $\{C_i(p)\}$ from old sets
$\{C_i(p-1)\}$.

What is important in the above rule is that we only consider
two-nucleon collisions which occur successively starting with the
first collision of the incident proton as we explained above in time
order.  We give collision indices only to these two-nucleon collisions
and we call them indexed collisions.  Any other two-nucleon collisions
which are not induced by the first collision of the incident proton
have no collision index and are called non-indexed collisions.
Non-indexed collisions play no role in constructing the sets
$\{C_i(p)\}$. For example, the set $C_i(p)$ of a nucleon $i \neq 1$
remains to be a null set for any $p$ if this nucleon $i$ is not
involved at all in any indexed collision, even when it experiences
many non-indexed collisions.

Thus the set $C_i(p)$ is a set composed of all the indexed collisions
that have had influence on the $i$-th nucleon at the moment just after
the No.$p$ collision.
If the total number of collision indices contained in $C_i(p)$ is $N$,
the nucleon $i$ after the No.$p$ collision is regarded as having a
step number $N$.  Especially when an outcoming nucleon has a step
number $N$, this nucleon is defined to be due to $N$-step process.

In Fig.\ 1 we give two illustrative examples of multi-step process. In
both cases (i) and (ii) of this figure, the nucleon (a) is the
incident proton and collision points are labeled by collision indices
1, 2, $\dots$ instead of No.1, No.2, $\dots$. In the case of (i), both
of outcoming nucleons (b) and (c) carry the same set of collision
indices $\{1, 2, 3 \}$ and hence are three-step nucleons.  In the case
of (ii), the nucleon (b) carries a set of collision indices $\{1, 2, 3
\}$, and hence is a three-step nucleon, while the nucleon (c) carrying
a set of collision indices $\{1, 2, 3, 4 \}$ is a four-step nucleon.
We should note here that the three-step nucleons (b) and (c) in the
case of (i) are formed as the result of a different type of collision
process from the three-step nucleon (b) in the case of (ii).  In the
case of (i), the three-step nucleons are formed by the collision of a
two-step nucleon with a zero-step nucleon. On the other hand, in the
case of (ii), the three-step nucleon is produced by a collision of a
two-step nucleon with a one-step nucleon.  In general, the type of
collision process in the case of (i), namely the collision of a
two-step nucleon with a zero-step nucleon, is more frequent in
producing a three-step nucleon.

Recently Kawai and his collaborators have developed a semi-classical
distorted wave model of nucleon inelastic scattering to continuum
\cite{LUOKAWAI,KAWAIWEIDE,WATAKAWAI}.  In this model, the first and
second Born terms are shown to correspond to one-collision and
two-collision processes in the INC model and are called one-step and
two-step processes, respectively.  Our above definition of one-step
and two-step processes is clearly the same as that in this
semi-classical distorted wave model.  We further expect that our
$N$-step process higher than two-step process will be proved to
correspond to the $N$-th Born term if their semi-classical distorted
wave model is extended to higher Born terms.

\section{Comparison with data}

We have studied $^{12}$C(p, p$'$x) at $E_p=$ 200 and 90 MeV, and
$^{58}$Ni(p, p$'$x) at $E_p=$ 120 MeV.
The total number of events is 4,000 for $^{12}$C target case while
for $^{58}$Ni target case it is 14,000 for case-1 $\sigma_{NN}$
and 3,000 for case-2 and case-3 $\sigma_{NN}$.
As stated above, in the case
of $^{12}$C target, we have used only case-1 $\sigma_{NN}$. The
data we compare with our calculations are taken from Ref.\
\cite{CARBON} and Ref.\ \cite{NICKEL} for $^{12}$C and $^{58}$Ni
targets, respectively.  The range of the adopted impact parameter $b$
is $0 \leq b \leq 8$ fm in the case of $^{12}$C target, while $0 \leq
b \leq 10$ fm in the case of $^{58}$Ni target.  If the incident proton
comes out without making any two-nucleon collision, we regard that the
event should not be included into inelastic scattering events to
continuum. In the case of $^{58}$Ni target, we show in Fig.\ 2 the
$b$-dependence of the probability $P_{\rm n.col.}(b)$ that the
incident proton comes out without making any two-nucleon collision.
It is to be noted that the total reaction cross section $\sigma_R$
which we discuss later is calculated as
\begin{eqnarray}
  \sigma_R  =  \int_0^\infty 2 \pi b db \,
  \Bigl( 1 - P_{\rm n.col.}(b) \Bigr).
\label{REACTCROSS}
\end{eqnarray}

The calculation of the double differential cross section
$d^2\sigma/d\Omega dE_{p'}$ is made in the following way;
\begin{eqnarray}
&& {d^2\sigma \over d\Omega dE_{p'}} =
  \int_0^\infty 2 \pi b db
  {d^2 {\cal N}( {\bf p}, b) \over d\Omega dE_{p'}}   , \nonumber\\
&& {d^2 {\cal N}( {\bf p}, b) \over d\Omega dE_{p'}} d\Omega dE_{p'}
  = \biggl< \sum_{\scriptstyle i =\rm isolated\atop
                             \hfill\scriptstyle\rm protons}
 |\langle {\bf p} | \phi_{{\bf Z}_i} \rangle|^2 \biggr>_b d^3p ,
\label{ANGDIS}\\
&& |\langle {\bf p} | \phi_{{\bf Z}_i} \rangle|^2
  = \Bigl( {1 \over 2 \pi \hbar^2 \nu} \Bigr)^{3/2}
  \exp \Bigl[ - {1 \over 2 \hbar^2 \nu} ( {\bf p} - {\bf K}_i )^2
  \Bigr] \nonumber
\end{eqnarray}
where $d^3p = m p d\Omega dE_{p'}$, and $\langle\;\;\rangle_b$
stands for the average value over the events with impact
parameter $b$. In this formula, the outcoming
protons are expressed by Gaussian wave packets with momentum width
$\hbar \sqrt \nu$.  In the case of protons emitted during the
dynamical stage, we adopt this width, but in the case of evaporated
protons from a fragment of mass number $A_F$, we adopt narrower width
by $1/\sqrt {A_F}$, namely $\hbar \sqrt {\nu / A_F}$.  It is because
the center-of-mass motion of the fragment of mass number $A_F$ is
described by a Gaussian wave packet whose momentum width is $\hbar
\sqrt {A_F \nu}$, from which the standard deviation of momentum per
nucleon is given by $\hbar \sqrt {\nu / A_F}$.  In the actual
calculation of the angular distribution, we further make another
modification to $|\langle {\bf p} | \phi_{{\bf Z}_i} \rangle|^2$.
It is the cut of the high
momentum tail of $|\langle {\bf p} | \phi_{{\bf Z}_i} \rangle|^2$. The
reason of this tail truncation is because the very high momentum part
of the distribution $|\langle {\bf p} | \phi_{{\bf Z}_i} \rangle|^2$
has no physical justification.  The tail part which we truncate is
defined by the condition $( {\bf p} - {\bf K}_i )^2 / 2 \hbar^2 \nu
\geq 1.8^2$ for dynamical nucleons, and
$A_F( {\bf p} - {\bf K}_i )^2 / 2 \hbar^2 \nu\geq 1.8^2$
for evapolated nucleons.
The volume of this tail part is about 10 $\%$ of the total volume of
$|\langle {\bf p} | \phi_{{\bf Z}_i} \rangle|^2$.

In Fig.\ 3 we give, in the case of the $^{12}$C target, the comparison
of the calculated angular distribution $d^2\sigma/d\Omega dE_{p'}$
with the data for various values of outcoming proton energy
$E_{p'}$. We see that the reproduction of data is good.  It is to be
noted that good reproduction has been obtained without any adjustable
parameter.  In the case of $E_p=$ 90 MeV, the cross sections at large
angles for $E_{p'}=$ 55 and 45 MeV are under estimated. We think that
the event number 4,000 is still not enough to reproduce large angle
cross sections. From the study of $^{58}$Ni(p,p$'$x) shown in Fig.\ 6,
we can guess that we need to increase the event number to about
10,000.  In Figs.\ 4 and 5 we show the decomposition of the calculated
angular distribution into multi-step contributions for the cases of
$E_{p}$ = 200 and 90 MeV, respectively.  We see that for our present
region of $E_{p'} \agt 45$ MeV, the contribution of one-step process
is predominant over a wide angular range but yet at large angles we
can recognize some predominance of two-step process. Especially in the
case of highly inelastic scattering with $E_{p'}$ = 45 MeV for $E_{p}$
= 200 MeV, the two-step process is dominant over a fairly wide angular
range and is indispensable for the reproduction of data.

The comparison of the calculated angular distribution with the data in
the case of the $^{58}$Ni target is given in Fig.\ 6.  We see that the
reproduction of data is again good in view of the absence of any
adjustable parameter in our approach.  What is to be noted here is the
result that almost the same good reproduction of data is obtained for
all three choices of in-medium cross section $\sigma_{NN}$, case-1
$\sigma_{NN}$, case-2 $\sigma_{NN}$, and case-3 $\sigma_{NN}$.  To
state in more detail, case-1 $\sigma_{NN}$ gives slightly better
data-fitting for $E_{p'}$ = 100 MeV, while case-3 $\sigma_{NN}$ gives
somewhat better fitting for $E_{p'}$ = 40 MeV.

The decomposition into multi-step contributions of the calculated
angular distribution is given in Fig.\ 7. The decomposition of the
calculated angular distribution by case-2 $\sigma_{NN}$ is similar
to that by case-3 $\sigma_{NN}$. We see
that two-step contribution is dominant at some large angle region and
is indispensable for the reproduction of data.  What is to be noted is
the forward cross section in the case of highly inelastic scattering
with $E_{p'}$ = 40 MeV.  In this case the two-step contribution is
much larger than the one-step contribution in the forward cross
section in all the three cases of adopted in-medium $\sigma_{NN}$.  It
is because the quasi-free peak of the one-step process is located in
fairly large angle region in the case of highly inelastic scattering.
Another point to be noted is the contribution of three-step process in
large angle region. Fig.\ 7 shows that in both cases of
$\sigma_{NN}$ the data points above $\theta =$ 100$^\circ$ for
$E_{p'}$ = 100 MeV are reproduced mainly by three-step contributions.
Also for $E_{p'}$ = 60 MeV, the data points above $\theta =$
120$^\circ$ have large contributions from the three-step process.  We
clearly see that the two-step contribution is much more important in
this case of $^{58}$Ni target than in the case of $^{12}$C target,
which is of course quite natural.

In Fig.\ 7, we indicated with arrows ($\alpha$) the angle of the
quasi-free peak which is given by $\cos^{-1} \sqrt {E_{p'}/E_{p}}$.
The peak angles of the calculated one-step angular distributions are
however seen to be slightly shifted to forward direction.  One of
the reasons of this shift is the mean field effect.  If the collision
takes place at the point of the potential depth $V$, the angle of the
quasi-free peak is given by $\cos^{-1} \sqrt {(E_{p'}-V)/(E_{p}-V)}$.
When we take $V=-50$ MeV, the shifts to forward direction of the
angles of the quasi-free peak at $E_{p'}=$ 100, 60, and 40 MeV are
$4^\circ$, $9^\circ$, and $12^\circ$, respectively. We show in Fig.\ 7
by arrows ($\beta$) these shifted angles of the quasi-free peak
obtained with $V=-50$ MeV. We see that these shifted angles are now
much closer to the peak angles of the calculated one-step angular
distributions. Another possible reason of the angle shift is the
refraction effect in the surface region of the target which makes the
path of one-step nucleon bend to forward direction.

Recently Shinohara et al.\cite{KAWAI} studied this reaction,
$^{58}$Ni(p, p$'$x) at $E_p=$ 120 MeV, by the use of a semi-classical
distorted wave model of nucleon inelastic scattering to continuum
\cite{LUOKAWAI,KAWAIWEIDE,WATAKAWAI}.  Our angular distribution due to
the one-step process is very similar to that of Ref.\ \cite{KAWAI},
while the two-step contribution of Ref.\ \cite{KAWAI} is fairly larger
than that of our calculation.  Especially the bigger contribution of
the two-step process than the one-step process seen in the case of
$E_{p'}$ = 60 MeV in Ref.\ \cite{KAWAI} is in disagreement with our
result.

\section{Dependence on in-medium N-N cross section}

We have seen in the previous section that the calculated angular
distribution $d^2\sigma/d\Omega dE_{p'}$ of $^{58}$Ni(p, p$'$x) at
$E_p=$ 120 MeV for $E_{p'} \agt 40$ MeV is insensitive to the choice
of different in-medium cross section $\sigma_{NN}$. We investigate
here the reason of it.

In Table II we show the calculated reaction cross section $\sigma_R$
(Eq.\ (\ref{REACTCROSS})) by the use of case-1 $\sigma_{NN}$, case-2
$\sigma_{NN}$, and case-3 $\sigma_{NN}$.  We see clear difference
between the values of $\sigma_R$'s obtained by different
$\sigma_{NN}$.  Reflecting the relation $(\sigma_{NN})_{\text{case-1}}
< (\sigma_{NN})_{\text{case-2}} < (\sigma_{NN})_{\text{case-3}}$, the
resulting $\sigma_R$ satisfy $(\sigma_R)_{\text{case-1}} <
(\sigma_R)_{\text{case-2}} < (\sigma_R)_{\text{case-3}}$.  As is
reflected in Eq.\ (\ref{REACTCROSS}), the magnitude of $\sigma_R$ is
determined only by the magnitude of the cross section of the first
collision of the incident proton.  Namely the magnitude of $\sigma_R$
does not reflect any information of the later reaction process after
the first collision of the incident proton.  On the other hand, the
angular distribution for a fixed value of $E_{p'}$ reflects not only
first collision of the incident proton but also the reaction process
after the first collision.  Let us compare the reaction process due to
$(\sigma_{NN})_{\text{case-1}}$ with that due to
$(\sigma_{NN})_{\text{case-3}}$.  Since
$(\sigma_{NN})_{\text{case-3}}$ is larger than
$(\sigma_{NN})_{\text{case-1}}$, the probability of the first
collision by $(\sigma_{NN})_{\text{case-3}}$ is higher than that by
$(\sigma_{NN})_{\text{case-1}}$.  But at the same time, the
probability of the second and further collisions by
$(\sigma_{NN})_{\text{case-3}}$ is also higher than those by
$(\sigma_{NN})_{\text{case-1}}$.  Therefore the energy of the
outcoming protons are much more damped in the average for
$(\sigma_{NN})_{\text{case-3}}$ than for
$(\sigma_{NN})_{\text{case-1}}$.  Thus the cross section
$d\sigma/dE_{p'}$ is not necessarily larger for
$(\sigma_{NN})_{\text{case-3}}$ than for
$(\sigma_{NN})_{\text{case-1}}$, at the $E_{p'}$ value where the
one-step process makes large contributions.  This compensation
mechanism between the first-collision probability and the energy
damping due to multiple collisions explains why the calculated angular
distribution at fixed value of $E_{p'}$ is almost independent of the
in-medium N-N cross section within the present three kinds of
$\sigma_{NN}$.

In Table II we have also shown the cross sections
$\sigma_{\text{1-step}}$ of outcoming protons due to the one-step
process, for three cases of $\sigma_{NN}$.  We see that
$\sigma_{\text{1-step}}$ is smaller for larger $\sigma_{NN}$.  This
result supports our above argument.  If $\sigma_{NN}$ is large, the
nucleon after the first collision will make further collisions rather
than escaping from the target without making further collisions.  If
this effect is bigger than the effect of the large cross section of
the first collision, $\sigma_{\text{1-step}}$ becomes smaller for
larger $\sigma_{NN}$. Table II shows that it is actually the case.

The cross section $\sigma_{\rm dyn}$ in Table II is the cross section
of dynamical protons, namely protons emitted during dynamical stage.
We see that $\sigma_{\rm dyn}$ is not so much dependent on the
magnitude of $\sigma_{NN}$. It is because the cross sections of
protons due to higher multi-step processes tend to become larger for
larger $\sigma_{NN}$, although $\sigma_{\text{1-step}}$ is smaller for
larger $\sigma_{NN}$.

In Table II we also give the calculated values of the cross sections
$\sigma_{\rm evap}$ of evaporated protons.  Quite naturally
$\sigma_{\rm evap}$ is larger for larger $\sigma_{NN}$.  The cross
section $\sigma_{p'}$ in Table II is the total cross section of all
the outcoming protons, namely $\sigma_{p'} = \sigma_{\rm evap} +
\sigma_{\rm dyn}$.  The dependence of $\sigma_{p'}$ on $\sigma_{NN}$
is due to that of $\sigma_{\rm evap}$.

In the approach with the semi-classical distorted wave model of Refs.\
\cite{KAWAI,LUOKAWAI,KAWAIWEIDE,WATAKAWAI}, the imaginary part
$W_{\rm opt}$ of the optical potential and the in-medium cross section
$\sigma_{NN}$ are important ingredients of the model.  These $W_{\rm
opt}$ and $\sigma_{NN}$ are mutually intimately related and hence the
choice of these quantities should be made consistently. It means that
the change of the magnitude of $\sigma_{NN}$ needs to be made together
with the consistent change of $W_{\rm opt}$.  On the other hand, in
the case of our AMD approach, we need not to introduce $W_{\rm opt}$
and the role of $W_{\rm opt}$ is described automatically in the
many-body dynamics which includes the two-nucleon collision process.
Therefore when we change the $\sigma_{NN}$ value in the AMD approach,
it implies that such effects that are described by $W_{\rm opt}$ are
changed consistently in an automatic way.  This merit of the AMD
approach is very advantageous in our present discussion of the
dependence of $d^2\sigma/d\Omega dE_{p'}$ on $\sigma_{NN}$. It is,
however, to be noted that in principle there exists mutual relation
between $\sigma_{NN}$ and the effective nuclear force. Hence the
change of $\sigma_{NN}$ is to be correlated with the change of the
effective nuclear force.  In our present study we study the dependence
of $d^2\sigma/d\Omega dE_{p'}$ on $\sigma_{NN}$ by assuming that the
adoption of the Gogny force as the effective nuclear force is more
reliable than the choice of $\sigma_{NN}$.

We have seen that the reaction cross section $\sigma_R$ depends rather
strongly on $\sigma_{NN}$.  Therefore the comparison of the calculated
$\sigma_R$ with data is expected to give us important information on
$\sigma_{NN}$. We here should recall that $\sigma_R$ is determined by
the cross section of the very first collision of the incident
proton. Hence $\sigma_{NN}$ which we expect to extract from $\sigma_R$
data is that in the region of the incident energy $E_p$.  To the
knowledge of the present authors, the $\sigma_R$ data of the $^{58}$Ni
+ p system at $E_p = 120$ MeV are not available.  However, as we
discuss below, we have several reasons to expect that
$\sigma_R(^{58}{\rm Ni},120\ {\rm MeV})$ lies in the region of $650
\sim 800$ mb. In Ref.\ \cite{NADA} in which $\sigma_R$ of $^{40}$Ca +
p, $^{90}$Zr + p, and $^{208}$Pb + p systems are experimentally
studied in the wide energy range up to 200 MeV, we see that the
approximate target-mass-number ($A_T$) dependence and the approximate
energy dependence of $\sigma_R$ in all these three systems are
$\sigma_R \propto A_T^{2/3}$ and $d \sigma_R / d \ln (E_p/{\rm MeV})
\approx -170$ mb, respectively, in the region of 50 MeV $\alt E_p
\alt$ 150 MeV.  When we apply this empirical rule of $A_T$ dependence
to the data $\sigma_R(^{40}{\rm Ca},120\ {\rm MeV})$ and
$\sigma_R(^{90}{\rm Zr},120\ {\rm MeV})$ given in Fig.\ 12 of Ref.\
\cite{NADA}, we get $\sigma_R(^{58}{\rm Ni},120\ {\rm MeV}) \approx $
750 mb. According to Ref.\ \cite{BAUHOFF}, observed
$\sigma_R(^{58}{\rm Ni})$ at $E_p$ = 60.8 MeV is 807 mb. When we apply
the above empirical rule of energy dependence to this, we obtain
$\sigma_R(^{58}{\rm Ni},120\ {\rm MeV})$ $\approx$ 690 mb.
Furthermore in Ref.\ \cite{BAUHOFF} it is reported that
$\sigma_R(^{59}{\rm Co})$ = 780 mb at $E_p$ = 98.5 MeV and
$\sigma_R({\rm Cu})$ = 751 mb at $E_p$ = 113 MeV.  All the above data
and estimations support that $\sigma_R(^{58}{\rm Ni})$ at $E_p$ = 120
MeV is in the region of 650 $\sim$ 800 mb. When we accept this range
of value for $\sigma_R(^{58}{\rm Ni})$ at $E_p$ = 120 MeV, we can
extract the conclusion that case-1 $\sigma_{NN}$ is most plausible
in the region of $E_p$ = 120 MeV among the three cases of
$\sigma_{NN}$, although case-1 $\sigma_{NN}$ gives still rather
large $\sigma_R$.

In Fig.\ 6 we see that in the case of $E_{p'}$ = 40 MeV the calculated
value of $d^2\sigma/d\Omega dE_{p'}$ is smallest in the case of the
case-3 $\sigma_{NN}$ than case-1 and case-2 $\sigma_{NN}$.  This
makes the reproduction of the observed angular distribution by the
case-3 $\sigma_{NN}$ slightly better than that by other $\sigma_{NN}$.
Since the contribution of the one-step process is largest in the angle
region around quasi-free peak, we can understand the situation as
follows: In order for the proton with $E_{p'} \approx$ 40 MeV after
the first collision to escape out of the target, this proton should
not make second collision.  If $\sigma_{NN}$ is large, the probability
to make the second collision is large and hence the value of
$d^2\sigma/d\Omega dE_{p'}$ due to one-step process becomes smaller.
This argument implies that the in-medium cross section $\sigma_{NN}$
in the low energy region around 40 MeV is better represented by the
case-3 $\sigma_{NN}$ than other case-1 and case-2 $\sigma_{NN}$.
Namely, $\sigma_{NN}$ in the low energy region around 40 MeV is
suggested to be closer to the free cross section to than case-1
and case-2 $\sigma_{NN}$.  But of course much more investigations are
needed to have conclusive justification about this point.

\section{Summary}

Proton inelastic scatterings to continuum $^{12}$C(p, p$'$x) at $E_p=$
200 MeV and 90 MeV and $^{58}$Ni(p, p$'$x) at $E_p=$ 120 MeV have been
studied with AMD (antisymmetrized molecular dynamics) which has been
successfully utilized for the study of heavy-ion collisions.  Angular
distributions $d^2\sigma/d\Omega dE_{p'}$ for various $E_{p'}$
energies have been shown to be reproduced well without any adjustable
parameter.  It shows the reliability and usefulness of AMD in
describing light-ion reactions.

Decomposition of the calculated angular distributions into multi-step
contributions has been made and two-step contributions have been found
to be dominant in some large angle region and to be indispensable for
the reproduction of data.  In the case of highly inelastic scattering
with $E_{p'}$ = 40 MeV of $^{58}$Ni(p, p$'$x) at $E_p=$ 120 MeV,
two-step contributions have been found to overwhelm one-step
contributions in forward cross sections. Furthermore in the
$^{58}$Ni(p, p$'$x) case, even three-step processes have been found to
make dominant contributions in the reproduction of data in the large
angle region $\theta \agt 120^\circ$ for $E_{p'}$ = 100 MeV.

We have studied the dependence of the calculated angular distributions
$d^2\sigma/d\Omega dE_{p'}$ for various $E_{p'}$ energies on the
in-medium N-N cross section $\sigma_{NN}$ in the case of $^{58}$Ni(p,
p$'$x) at $E_p=$ 120 MeV, and have found that the dependence is very
weak. The reason of this insensitivity of $d^2\sigma/d\Omega dE_{p'}$
to $\sigma_{NN}$ has been clarified by studying also the
$\sigma_{NN}$-dependence of various kinds of cross sections such as
the reaction cross section $\sigma_R$, the one-step-proton cross
section $\sigma_{\text{1-step}}$, and the evaporated-proton cross section
$\sigma_{\rm evap}$. Unlike $d^2\sigma/d\Omega dE_{p'}$, these quantities,
$\sigma_R$, $\sigma_{\text{1-step}}$, and $\sigma_{\rm evap}$, have been shown
to
be sensitive to $\sigma_{NN}$.  For the choice of larger
$\sigma_{NN}$, $\sigma_R$ and $\sigma_{\rm evap}$ become larger, while
$\sigma_{\text{1-step}}$ becomes smaller for larger $\sigma_{NN}$.

\acknowledgements

The authors would like to thank Professors M. Kawai and Y. Watanabe
for helpful discussions.  Helpful comments of Professor D. Brink are
gratefully acknowledged.  They are also grateful to Professor
H. Sakaguchi for helpful comments on experimental data.  Their thanks
are also due to Professors A. A. Cowley and J. J. Lawrie for kindly
sending us detailed data values of $^{12}$C(p, p$'$x) and $^{58}$Ni(p,
p$'$x).  They also would like to thank Miss S. Furihata for
discussions about the calculation of the step numbers in the AMD study
of $^{12}$C(n, px) reactions. Numerical calculations were made mainly
by using the Fujitsu VPP500 of RIKEN.

\widetext
\begin{figure}
\epsfxsize=\textwidth\epsfbox{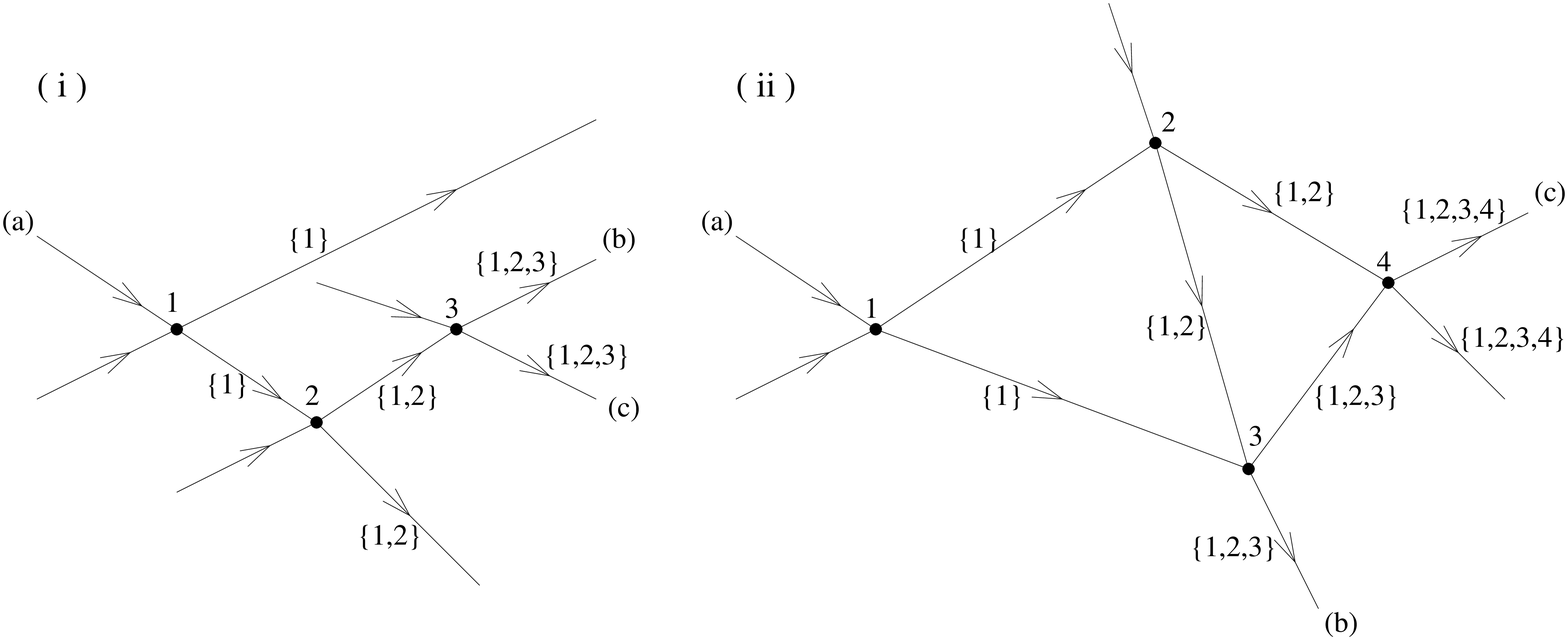}\vspace{3ex}
\caption{
Illustrative examples of multi-step process. Nucleon (a) is the
incident proton and collision points are labeled by collision indices
1, 2, $\dots$ instead of No.1, No.2, $\dots$. In the figure (i), the
outcoming nucleons (b) and (c) carry the same set of collision indices
$\{1, 2, 3 \}$ and hence are three-step nucleons.  In the figure (ii),
the exit nucleon (b) is a three-step nucleon since it carries a set
$\{1, 2, 3 \}$ composed of three collision indices, while the exit
nucleon (c) carrying a set $\{1, 2, 3, 4 \}$ is a four-step nucleon.  }
\end{figure}
\narrowtext

\begin{figure}
\epsfxsize=0.5\textwidth\epsfbox{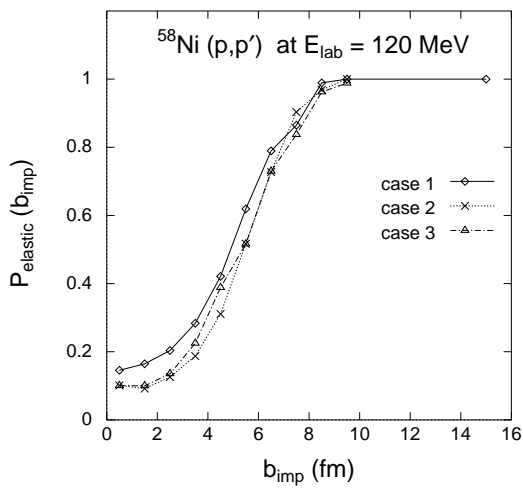}
\caption{
Impact-parameter (b) dependence of the probability $P_{\rm n.col.}$ that
the incident proton comes out without making any two-nucleon
collision. Calculations are made for $^{58}$Ni(p, p$'$x) at $E_p=$ 120
MeV by the use of three kinds of in-medium N-N cross section
$\sigma_{NN}$ explained in the text.}
\end{figure}

\widetext
\begin{figure}
\epsfxsize=\textwidth\epsfbox{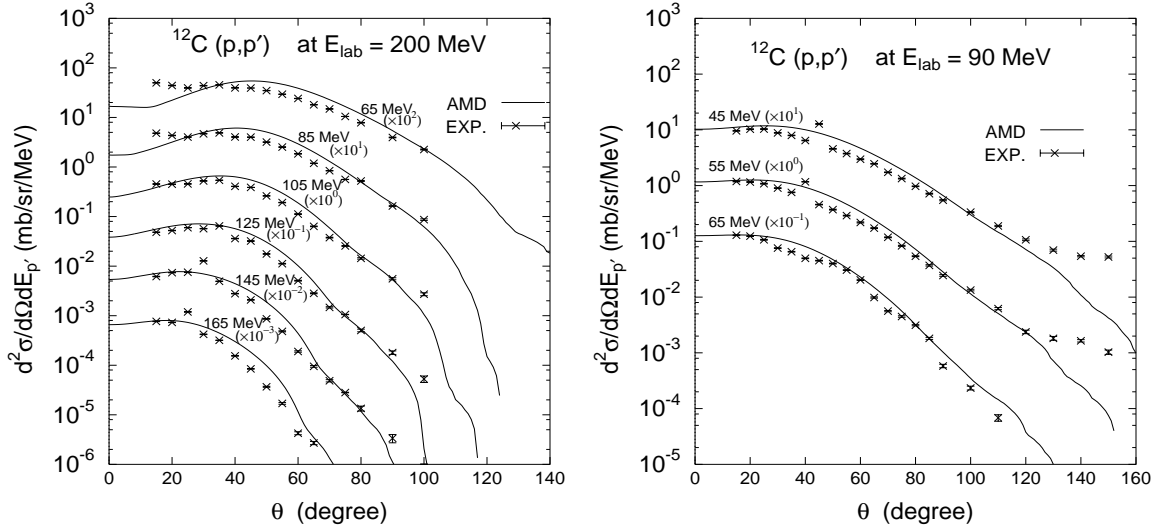}
\caption{
Comparison of calculated angular distribution $d^2\sigma/d\Omega
dE_{p'}$ of $^{12}$C(p, p$'$x) with data for various $E_{p'}$
energies.  Comparisons are made in two cases of incident proton
energies $E_p=$ 200 MeV and 90 MeV.}
\end{figure}
\narrowtext

\widetext
\begin{figure}
\epsfxsize=\textwidth\epsfbox{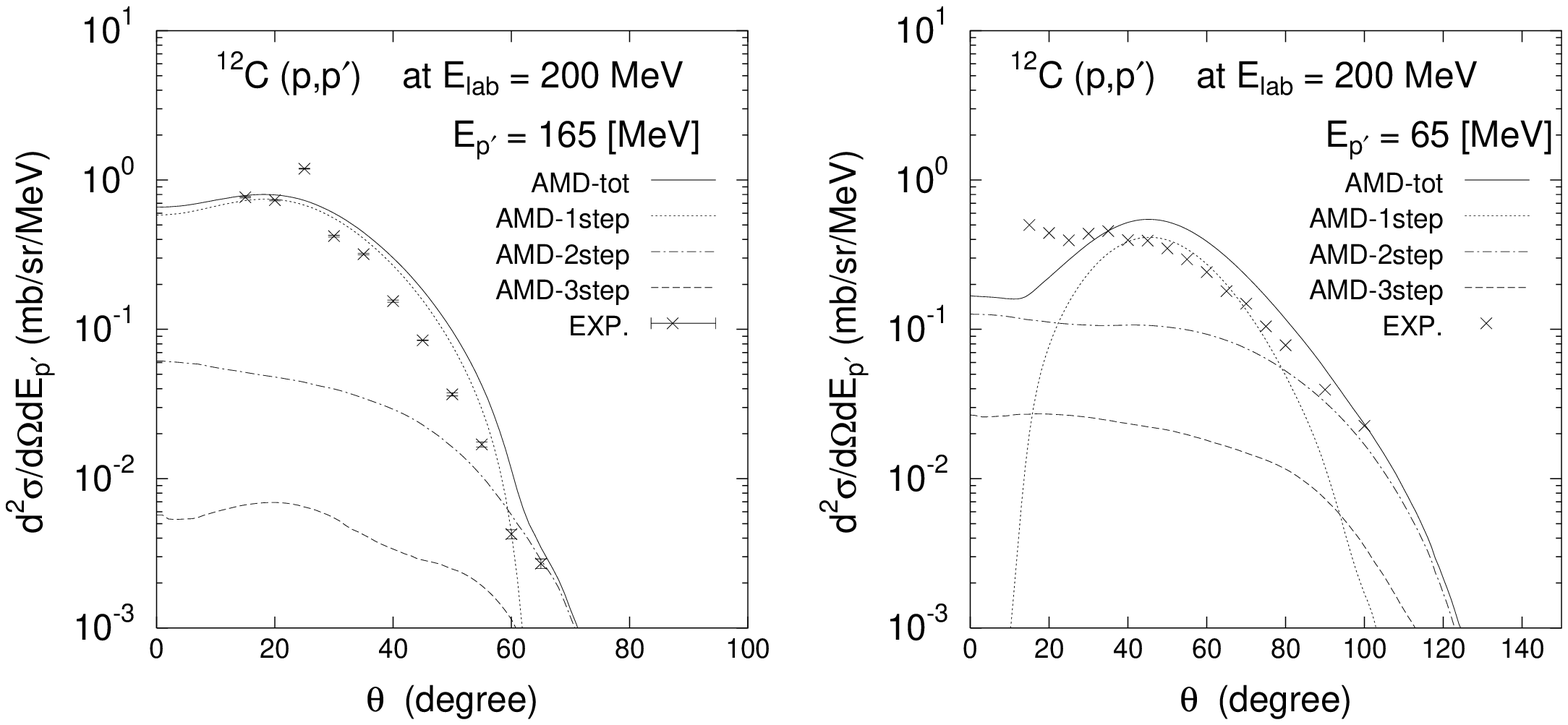}
\caption{
Decomposition of the calculated angular distributions into multi-step
contributions in the case of $^{12}$C(p, p$'$x) at $E_p=$ 200 MeV.  }
\end{figure}
\narrowtext

\widetext
\begin{figure}
\epsfxsize=\textwidth\epsfbox{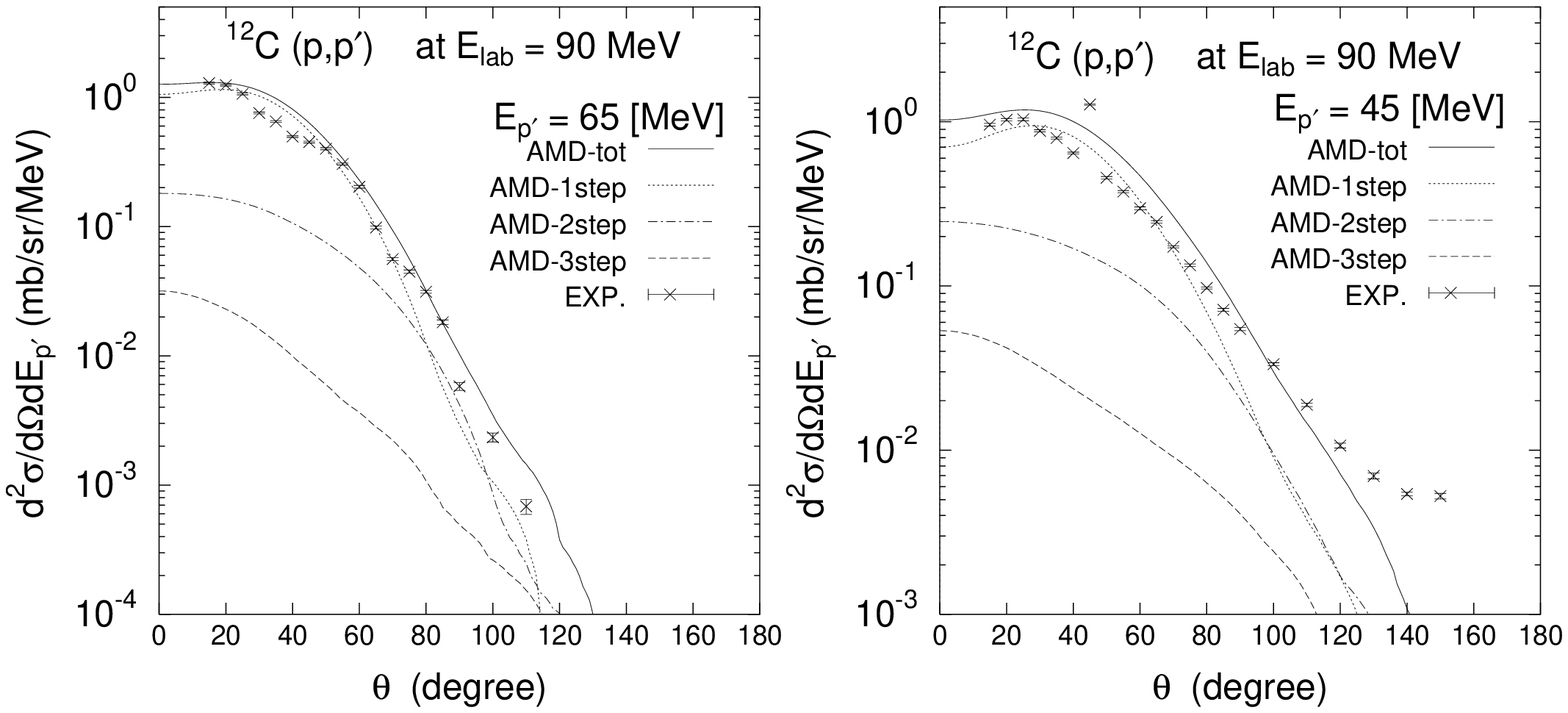}
\caption{
Decomposition of the calculated angular distributions into multi-step
contributions in the case of $^{12}$C(p, p$'$x) at $E_p=$ 90 MeV.  }
\end{figure}
\narrowtext

\begin{figure}
\epsfxsize=0.5\textwidth\epsfbox{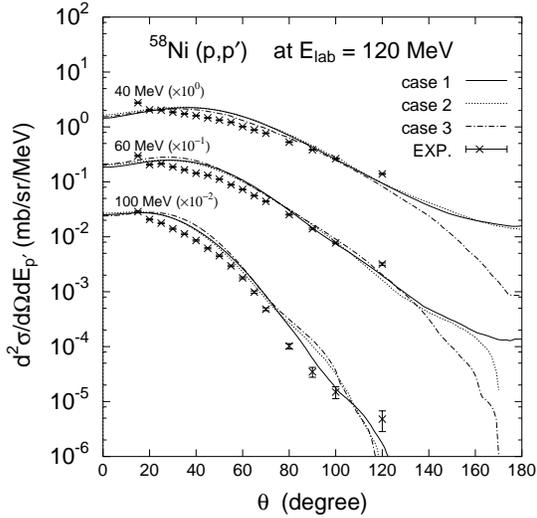}
\caption{
Comparison with data of calculated angular distribution
$d^2\sigma/d\Omega dE_{p'}$ of $^{58}$Ni(p, p$'$x) with incident
proton energy $E_p=$ 120 MeV. Comparisons are made for three kinds of
calculations performed by adopting three different in-medium N-N cross
sections; case-1 $\sigma_{NN}$, case-2 $\sigma_{NN}$, and case-3
$\sigma_{NN}$.}
\end{figure}

\widetext
\begin{figure}
\epsfxsize=\textwidth\epsfbox{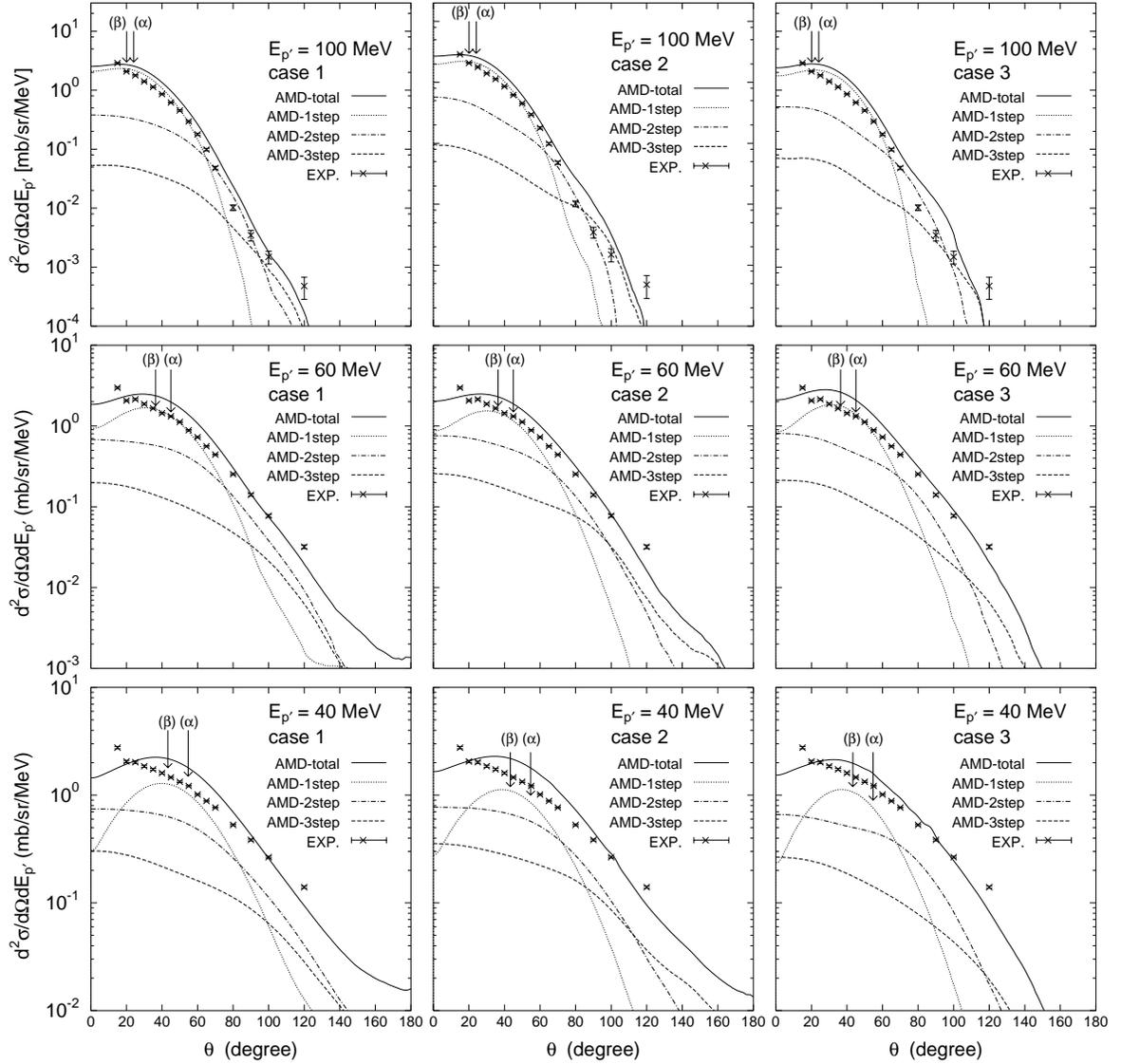}
\caption{
Decomposition of the calculated angular distributions of $^{58}$Ni(p,
p$'$x) at $E_p=$ 120 MeV into multi-step contributions. The angles
indicated with arrows ($\alpha$) and ($\beta$) are quasi-free
scattering angles given by $\cos^{-1} \protect\sqrt { E_{p'}/E_{p} }$
and $\cos^{-1}
\protect\sqrt {(E_{p'}-V)/(E_{p}-V)}$ with $V=-50$ MeV, respectively.
}
\end{figure}
\narrowtext

\begin{table}
\caption{
Values of parameters $A_1(E)$, $B_1(E)$, and $B_2(E)$ in mb/sr as
functions of the incident particle laboratory energy $E$, used in the
parametrization of the angular distribution of $\sigma_{pn}$ of the
case-3 $\sigma_{NN}$ which is taken from Ref.\ \protect\cite{BERTINI}.
See Eq.\ (\protect\ref{SIGMAANGZ}). }
\begin{tabular}{c|ddd}
$E$ (MeV) & $A_1(E)$ & $B_1(E)$ & $B_2(E)$ \\
\hline
  40 & 12.0 & 7.0 & 7.0 \\
  80 & 5.2 & 8.1 & 8.3 \\
 120 & 3.3 & 6.6 & 9.0 \\
 160 & 2.3 & 3.9 & 7.7 \\
 200 & 2.0 & 3.6 & 6.5 \\
 240 & 1.9 & 3.6 & 6.2 \\
 280 & 1.8 & 3.6 & 6.0 \\
\end{tabular}
\end{table}

\begin{table}
\caption{
Dependence of various kinds of cross sections on the different choices
of in-medium N-N cross section $\sigma_{NN}$, case-1 $\sim$ case-3.
The notations $\sigma_R$, $\sigma_{\text{1-step}}$, $\sigma_{\rm dyn}$,
$\sigma_{p^\prime}$, and $\sigma_{\rm evap}$stand for the reaction cross
section, the one-step-proton cross section, the dynamical-proton cross
section, the total-proton cross section, and the evaporated-proton
cross section, respectively.  Units are in mb.}
\begin{tabular}{l|rrr}
& {case-1} & {case-2} & {case-3} \\ \hline
 a )  $\sigma_R$ & 839 & 965 & 973 \\
 b )  $\sigma_{\text{1-step}} $& 589 &  518 & 489 \\
 c )  $\sigma_{\rm dyn} $ & 1027 & 1079 & 994 \\
 d )  $\sigma_{p^\prime}$ & 1749 & 2259 & 2308 \\
 e )  $\sigma_{\rm evap}$ & 722 & 1180 & 1314 \\
\end{tabular}
\end{table}

\end{document}